\documentclass[reprint]{revtex4-2}

\usepackage{bm}
\usepackage{graphicx}
\usepackage{amsmath}
\usepackage{amsbsy}
\usepackage{amstext}
\usepackage{amssymb}
\usepackage{units}
\usepackage{url}
\usepackage[colorlinks, citecolor=blue, urlcolor=blue, linkcolor=red,breaklinks]{hyperref}

\begin{document}
	
	\title{A general formulation for the magnetic oscillations in two dimensional systems}
	
	\author{F. Escudero}
	\author{J. S. Ardenghi}
	\author{P. Jasen}
	\affiliation{Departamento de F\'{i}sica, Universidad Nacional del Sur, Av. Alem 1253, B8000CPB Bah\'{i}a Blanca, Argentina\\
		Instituto de F\'{i}sica del Sur, Conicet, \\
		Av. Alem 1253, B8000CPB Bah\'{i}a Blanca, Argentina}
	
	\begin{abstract}
		We develop a general formalism for the magnetic oscillations (MO) in two dimensional (2D) systems. We consider general 2D Landau levels, which may depend on other variable or indices, besides the perpendicular magnetic field. In the ground state, we obtain expressions for the MO phase and amplitude. From this we use a Fourier expansion to write the MO, with the first term being a sawtooth oscillation. We also consider the effects of finite temperature, impurities or lattice imperfections, assuming a general broadening of the Landau levels. We develop two methods for describing these damping effects in the MO. One in terms of the occupancy of the Landau levels, the other in terms of reduction factors, which results in a generalization of the Lifshits-Kosevich (LK) formula. We show that the first approach is particularly useful at very low damping, when only the states close to the Fermi energy are excited. In contrast, the LK formula may be more convenient at higher damping, when only few terms are needed in its harmonic expansion. We compare different damping situations, showing how the MO are broadened in each case. The general formulation presented allows to relate the properties of the MO with those of the 2D systems.
	\end{abstract}
	
	\maketitle
	
	\section{Introduction}\label{sec:1}
	
	One of the most important consequences of the Landau quantization are the oscillations of the thermodynamic potentials. For instance, the de Haas-van Alphen effect is the oscillation of the magnetization as a function of the magnetic field \cite{de1930oscillations}. This phenomenon was first observed experimentally in bismuth in 1930, and theoretically explained by Landau \cite{Landau1930}, but for many years it was considered to be just a curiosity. It was later shown that, by analysing the magnetic oscillations (MO) frequencies, one could obtain information about the Fermi surface of the material \cite{Onsager1952}. In fact, the MO profile depends on the Landau levels (LL) of the system. In ideal conditions, any distinctive feature of the LL should be observed in the properties of the corresponding MO \cite{Escudero2018,Escudero2017a}. This is particularly revealing in two dimensional (2D) materials, which have many unique properties not seen in the conventional 2D electron gas \cite{Lin2016,Mas-Balleste2011,Gupta2015}.
	
	For example, in graphene-like systems \cite{Novoselov2005}, at low energy the LL are relativistic and thus not equidistant \cite{Neto2009,Goerbig2011}. Moreover, whereas the hexagonal lattice of graphene is planar \cite{Geim2007}, in other materials such as silicene \cite{Zhuang2015,Zhao2016,Spencer2016,Houssa2015} or germanene \cite{Acun2015,Balendhran2014,Ezawa2015}, the hexagonal lattice is buckled. Hence the energy bands can be split by applying a potential difference between them \cite{Drummond2012,Ni2011,Yan2015}. The energy of these systems can also depend strongly on the spin-orbit interaction (SOI) \cite{Liu2011}, which in turn can make them a topological insulator \cite{Kou2017,Tahir2013,Huang2016}. When the SOI is strong enough, is possible to observe the quantum spin Hall effect \cite{Hsu2017,Wang2016}. The LL are also modified when other external fields are applied, such as parallel electric field, or when the lattice itself is deformed \cite{Goerbig2011,Aghaei2015}. In addition, the LL may depend on the spin or valley properties of the 2D system \cite{Tahir2013,Schaibley2016,Young2012}, which can also influence the electronic and magnetic properties \cite{McCann2006,Ardenghi2015}. Thus, there can be a rich manipulation of the LL in 2D materials, all of which should be reflected in the corresponding MO \cite{Wilde2005}.
	
	Damping effects, due to temperature or impurity scattering, tend to broaden and reduce the MO amplitude \cite{Shoenberg1984a,Escudero2017b,Escudero2018a}. Besides the need for relatively strong magnetic fields, this damping of the MO is the reason why they are difficult to observe \cite{Escudero2019}. The damping effects are usually considered as reduction factors in the MO amplitude, which results in the known Lifshits-Kosevich (LK) formula \cite{lifshitz1956theory}. This formula is particularly useful in the case of relatively high damping, where the infinite series that gives the LK formula can be exactly solved \cite{Shoenberg1984}, or well approximated with the first few harmonic terms. That is commonly the case in 3D metals, where the oscillations can be treated as simple harmonic. Nevertheless, the higher the damping, the more details are lost in the MO. In 2D materials, this loss can be significant, as many properties of the system, such as the interplay between valley and spin, would no longer be observed in the MO. 
	Another approach to treat the damping of the MO, recently developed in pristine graphene and 2D materials \cite{Escudero2018a,Escudero2019}, considers the temperature effect by its modification to the MO in the ground state. At zero impurities, this modification is given by Fermi-Dirac distribution functions, and thus they represent the damping in the MO due to the change in the occupancy of the LL. This proved to be useful at very low temperatures, when only the states nearby the Fermi energy are excited and the damping is low enough to observe the MO fine details.
	
	Despite the recent advances in the study of the MO in 2D materials
	\cite{Kishigi2002,Zimbovskaya2007,Wang2009,Nasir2009,Qiong-Gui2010,Lukyanchuk2011,Tabert2014,Hese2014}, the analysis usually considers special cases, such as assuming only a perpendicular magnetic or electric field \cite{Zhang2010,Fu2011,Tabert2015,Fortin2015,Fortin2017,Becker2019}. There is no general formulation of the MO, in case the LL depend on other variables or indices.
	When damping effects are considered, they are usually described using the LK formula, assuming a Lorentz distribution for the LL broadening
	\cite{1952a}. However, there is no general consensus on how is the disorder in 2D materials, in the presence of a magnetic field \cite{Ando1982,Sharapov2004}. Depending on the situation, one impurity distribution could be more suitable than another \cite{Williamson1964,Fang1988,Gammag2008,Yang2010,Funk2015}. Thus, it is important to have a detailed analysis of how different broadening distributions modify the MO. In fact, careful measurements of the MO can be one of the most reliable methods to obtain information about the LL broadening \cite{Potts1996}.
	
	Motivated by this, in this paper we developed a general formulation for the magnetic oscillations in 2D systems, considering a general expression for the LL, and taking into account damping effects such as finite temperature, impurities or other lattice imperfections. We have organized this work as follow: in section \ref{sec:2} we define the general LL considered, assuming that they may depend in any variables and indices, and we develop a general formulation to obtain the corresponding ground state MO. We obtain expressions for the sawtooth oscillation amplitude $A$ and the oscillating phase $\psi$. From these we use a Fourier expansion which allow us to write the MO in terms of $A$ and $\psi$. In section \ref{sec:3} we consider the damping effects in the MO. We model the effect of impurities or lattice imperfections by considering a general density of states that broadens each LL. From this we obtain an expression for the damping effects in the MO, which depends on the occupancy of each LL. We then show that, if the broadening is the same for all LL, the damping effects can also be described by reduction factors, which leads to a generalization of the LK formula. As an example we consider the MO in graphene-like systems, comparing both approaches for different damping cases. Finally, our conclusions follow in section \ref{sec:4}.
	
	\section{Ground state magnetic oscillations}\label{sec:2}
	
	We consider a 2D system under a perpendicular magnetic field $B$,
	with discrete LL $\varepsilon_{n;\gamma}\left(B;X\right)$,
	where $n$ is the LL index. We denote with $\gamma$ all
	the other indices with which the energy levels may depend, such as
	the spin $s$, the valley $\eta$, or the wave vector $k$. With $X$
	we denote all the other variables (besides $B$) with which $\varepsilon_{n;\gamma}$
	may also depend, such as the perpendicular (parallel) electric field
	$\ensuremath{E_{\bot}\left(E_{\parallel}\right)}$ and/or the parallel
	magnetic field $B_{\parallel}$.
	It is worth emphasizing that in all cases there is a perpendicular
	magnetic field $B$ which causes the discrete $n$ LL, which is why
	we separate them. The other indices $\gamma$ and variables $X$ may or
	not be present, depending on the specific case studied. In
	general, each energy level has a degeneracy $\varrho$, which could
	be the degeneracy $D=\mathcal{A}Be/h$ (where $\mathcal{A}$ is the 2D system area), in case the
	energy levels do not depend on the wave vector, and/or any additional
	degeneracy of the system, such as the spin or valley degeneracy; the
	specific value of $\varrho$ depends on the particular system. We shall set the zero energy to be between the valence band (VB) and the conduction band (CB).
	
	Introducing the decreasing sorting index $m$ for the CB
	energy levels $\varepsilon_{m}$, for a constant chemical potential $\mu>0$, the ground state grand potential $\Omega^{0}$ is
	\begin{equation}
		\Omega^{0}=\Omega_{\text{VB}}^{0}+\sum_{m=1}^{f}\varrho\left(\varepsilon_{m}-\mu\right),\label{Omega}
	\end{equation}
	where $\Omega_{\text{VB}}^{0}$ is the valence band contribution and $f$
	is the last position such that $\varepsilon_{f}\leq\mu<\varepsilon_{f+1}$.
	The oscillation in $\Omega^{0}$
	is produced whenever $f$ changes, and hence only the CB contributes
	to the quantum oscillations. Therefore, since we are interested in
	the MO, we will omit the VB and work only with the CB grand potential.
	The magnetization is $M^{0}=-\mathcal{A}^{-1}\left(\partial\Omega^{0}/\partial B\right)_{\mu}$,
	where $\mathcal{A}$ is the 2D system area. For the CB
	we get
	\begin{equation}
		M_{\text{CB}}^{0}=-\frac{\partial\varrho}{\partial B}\frac{\Omega_{\text{CB}}^{0}}{\mathcal{A\varrho}}-\sum_{m=1}^{f}\frac{\varrho}{\mathcal{A}}\frac{\partial\varepsilon_{m}}{\partial B},\label{M0}
	\end{equation}
	where $\Omega_{\text{CB}}^{0}=\sum_{m=1}^{f}\varrho\left(\varepsilon_{m}-\mu\right)$.
	Given that $\Omega^{0}$ is always continuous, when $f$ changes to
	$f+1$ we have
	\begin{equation}
		\Delta M^{0}=-\frac{\varrho}{\mathcal{A}}\frac{\partial\varepsilon_{f+1}}{\partial B}\left(B_{f}\right).\label{DeltaM2}
	\end{equation}
	where $B_{f}$ is the magnetic field when the change occur (determined
	by $\varepsilon_{f+1}\left(B_{f}\right)=\mu$). This last equation gives the discontinuity in the ground state magnetization when the energy levels cross the chemical potential. The minus sign just tell us that the magnetization
	decreases when the number of filled states increases, given our initial
	assumption $f\rightarrow f+1$. 
	
	We shall now separate the magnetization peaks by the $\gamma$ indices, as previously defined. For
	each $\gamma$ index, we consider the magnetization peaks
	due to the change of the $n$ LL index. To see how this works, consider in general
	the peak that occurs when $\varepsilon_{n;\gamma}=\mu$, for each
	$\gamma$ index. Then we can solve $\varepsilon_{n;\gamma}=\mu$
	for $n$, obtaining in general 
	\begin{equation}
		\varepsilon_{n;\gamma}=\mu\longrightarrow n=\psi_{\gamma}\left(\mu,B,X \right).\label{phase}
	\end{equation}
	The function $\psi_{\gamma}$ will depend on $\mu$ and on the energy
	levels, but in all cases $\psi_{\gamma}$ determines when the peaks
	occurs, namely when $\psi_{\gamma}=n$ with $n$ a positive integer. 
	
	For each index $\gamma$, in the small-field limit the oscillating part of $\Omega_\gamma$ and $M_\gamma$ can be written in a Fourier series (see appendix \ref{app:A}). For $M_\gamma$, the first and dominant term in the series is a sawtooth oscillation (SO). The SO amplitude $A_{\gamma}$ is the discontinuity in $M_\gamma$ given by equation (\ref{DeltaM2}). Thus, considering the function $\psi_{\gamma}$, we have
	\begin{equation}
		A_{\gamma}=-\frac{\varrho}{\mathcal{A}}\left.\frac{\partial\varepsilon_{n;\gamma}}{\partial B}\right|_{n=\psi_{\gamma}}.\label{Ans}
	\end{equation}
	Then, the SO associated with each $\gamma$ can be written as
	\begin{equation}
		M_{\text{SO},\gamma}^{0}=A_{\gamma}\sum_{p=1}^{\infty}\frac{1}{\pi p}\sin\left(2\pi p\psi_{\gamma}\right).\label{Mns}
	\end{equation}
	The next order (non-sawtooth) term in the MO can be easily obtained considering that $\Omega^0$ is a continuous function and $M_{\text{osc}}^0=-\mathcal{A}^{-1}\left(\partial\Omega_{\text{osc}}^0/\partial B\right)_{\mu}$. This implies that $\Omega_{\text{osc}}^0$ is of the form  $\Omega_{\text{osc}}^0=\sum_{p=1}^{\infty}C_{2}\cos\left(2\pi p\psi_{\gamma}\right)/\left(\pi p\right)^{2}$, where $C_2$ is a $p$ independent factor that satisfies $2C_{2}\left(\partial\psi_{\gamma}/\partial B\right)/\mathcal{A}=A_{\gamma}$. Therefore we obtain
	\begin{align}
		\Omega_{\text{osc}}^{0} = & \sum_{\gamma}\frac{\mathcal{A}A_{\gamma}}{2\psi_{\gamma}^{\left(1\right)}}\sum_{p=1}^{\infty}\frac{\cos\left(2\pi p\psi_{\gamma}\right)}{\left(\pi p\right)^{2}}\\
		M_{\text{osc}}^{0} = & \sum_{\gamma}\left\{ M_{\text{SO},\gamma}^{0}+\frac{1}{2\psi_{\gamma}^{\left(1\right)}}\left[\frac{A_{\gamma}}{\psi_{\gamma}^{\left(1\right)}}\psi_{\gamma}^{\left(2\right)}-\left(\frac{\partial A_{\gamma}}{\partial B}\right)\right]\right.\nonumber \\
		& \left.\times\sum_{p=1}^{\infty}\frac{\cos\left(2\pi p\psi_{\gamma}\right)}{\left(\pi p\right)^{2}}\right\},\label{Mosc final} 
	\end{align} 
	where we noted $\psi_{\gamma}^{\left(n\right)}=\left(\partial^{n}\psi_{\gamma}/\partial B^{n}\right)$,
	the $n$th derivative of $\psi_{\gamma}$ with respect to $B$. These harmonic expansions are in agreement with the small-field limit expressions obtained in classical \cite{Shoenberg1984} and graphene-like \cite{Sharapov2004,Tabert2015} 2D systems. The expression (\ref{Mosc final})
	implies that, for any 2D energy levels $\varepsilon_{n;\gamma}$, the
	corresponding MO are determined by just two parameters:
	the dimensionless phase function $\psi_{\gamma}$ and the SO amplitude
	$A_{\gamma}$, which can be easily obtained from equations (\ref{phase})
	and (\ref{Ans}). The second term in $M_{\text{osc}}^{0}$ affects the
	MO by changing the curvature of the oscillations, but usually it can neglected if $\psi_{\gamma}\gg1$. To see this, consider that in general $\left(\partial A_{\gamma}/\partial B\right)$
	is very small if $A_{\gamma}$ depends on $B$, whereas $\psi_{\gamma}$
	decreases with $B$ if $\varepsilon_{n;\gamma}$ increases with $B$,
	as usual. Then the amplitude ratio between the cosine and sine series in
	equation (\ref{Mosc final}) is of order $\left|A_{\text{cos}}/A_{\text{sin}}\right|\sim1/\psi_{\gamma}$.
	Therefore at high occupancy the cosine series can be neglected and
	the MO can be approximated as a pure sawtooth oscillation. Hence, if
	$\psi_{\gamma}\gg1$ we simply have 
	\begin{equation}
		M_{\text{osc}}^{0}\simeq\sum_{\gamma}\frac{A_{\gamma}}{\pi}\arctan\left[\cot\left(\pi\psi_{\gamma}\right)\right].\label{Mosc high oc}
	\end{equation}
	It is worth noting that equation (\ref{Mosc high oc}) is valid
	at high occupancy, which generally
	requires relatively low $B$ and high $\mu$. 
	
	\section{Magnetic oscillations with damping}\label{sec:3}
	
	We shall consider the damping effects due to impurity scattering, or any other lattice imperfection, by the broadening of the energy levels, so each $\varepsilon_{n;\gamma}$
	will have a density of states (DOS) $\rho_{n;\gamma}\left(\varepsilon,\varepsilon_{n;\gamma},\Gamma_{n;\gamma}\right)$,
	where $\Gamma_{n;\gamma}$ is the parameter that defines the broadening.
	In general, depending on how is the DOS and the damping mechanism,
	the parameter $\Gamma_{n;\gamma}$ may depend on the same variables
	and indices as the the energy levels $\varepsilon_{n;\gamma}$. We will neglect the damping effects due to any many-body interaction, such as the scattering of electrons by phonons. This is a reasonable assumption, given the high Debye temperature of common 2D systems \cite{Balandin2011,Pop2012,Yang2014,Huang2015,Ren2016,Peng2016}, which are well above the very low temperature required to observe the MO. 
	Then the grand potential is
	\begin{equation}
		\Omega=-\frac{\varrho}{\beta}\int_{-\infty}^{\infty}\rho\ln\left[1+e^{\beta\left(\mu-\varepsilon\right)}\right]d\varepsilon,\label{Om}
	\end{equation}
	where $\beta=1/k_{\text{B}}T$, $\varrho$ is the degeneracy of each energy level
	(as in the ground state) and $\rho=\sum_{n;\gamma}\rho_{n;\gamma}$
	is the DOS. Notice that we separate the degeneracy $\varrho$ and the broadening $\rho$ of the LL; the total DOS would be $\varrho \rho$. Introducing again the energy sorting index $m$, as done in the previous section,
	we can write $\Omega=\sum_{m}\Omega_{m}$, with $\rho_{m}=\left(\varepsilon,\varepsilon_{m},\Gamma_{m}\right)$.
	Then, for each $m$ we have
	\begin{equation}
		\left(\frac{\partial\Omega_{m}}{\partial B}\right)_{\mu} =  \frac{\partial\varrho}{\partial B}\frac{\Omega_{m}}{\varrho} -\frac{\varrho}{\beta}\int_{-\infty}^{\infty}\frac{\partial\rho_{m}}{\partial B}\ln\left[1+e^{\beta\left(\mu-\varepsilon\right)}\right]d\varepsilon.\label{deltaOm}
	\end{equation}
	We shall assume the DOS $\rho_{m}$ to be symmetric around $\varepsilon_{m}$,
	such that $\partial\rho_{m}/\partial\varepsilon_{m}=-\left(\partial\rho_{m}/\partial\varepsilon\right)$.
	Considering
	that $\Gamma_{m}$ may also depend on $B$, integrating by parts we obtain 
	\begin{equation}
		\left(\frac{\partial\Omega_{m}}{\partial B}\right)_{\mu}=I_{1,m}+I_{2,m}+I_{3,m},\label{deltaOm2}
	\end{equation}
	where
	\begin{align}
		I_{1,m} = & -\frac{1}{\beta}\frac{\partial\varrho}{\partial B}\int_{-\infty}^{\infty}\rho_{m}\ln\left[1+e^{\beta\left(\mu-\varepsilon\right)}\right]d\varepsilon\label{I1}\\
		I_{2,m} = &  \varrho\frac{\partial\varepsilon_{m}}{\partial B}\int_{-\infty}^{\infty}\rho_{m}\frac{1}{1+e^{\beta\left(\varepsilon-\mu\right)}}d\varepsilon\label{I2}\\
		I_{3,m} = &  -\frac{\varrho}{\beta}\frac{\partial\Gamma_{m}}{\partial B}\int_{-\infty}^{\infty}\frac{\partial\rho_{m}}{\partial\Gamma_{m}}\ln\left[1+e^{\beta\left(\mu-\varepsilon\right)}\right]d\varepsilon.\label{I3}
	\end{align}
	From this, we will first consider the damping effects
	as a modification of the ground state MO, which depends on the occupancy
	of the LL. Then we will prove that this is equivalent to
	describe the damping effects with reduction factors, as in the LK formula.
	
	\subsection{Damping effects in terms of the occupancy of the Landau levels}\label{subsec:3.1}
	
	The idea is to describe the MO with damping in terms of its
	modification to the ground state MO. From equations (\ref{Omega}) and (\ref{deltaOm2}),
	summing over $m$ we can separate
	\begin{align}
		\left(\frac{\partial\Omega}{\partial B}\right)_{\mu} = & \left(\frac{\partial\Omega_{\text{VB}}}{\partial B}\right)_{\mu}+\left(\frac{\partial\Omega_{\text{CB}}^{0}}{\partial B}\right)_{\mu}+\sum_{m=1}^{\infty}I_{3,m}\nonumber \\
		& +\sum_{m=1}^{f}\left[I_{1,m}-\frac{\partial\varrho}{\partial B}\left(\varepsilon_{m}-\mu\right)\right]+\sum_{m=f+1}^{\infty}I_{1,m}\nonumber \\
		& +\sum_{m=1}^{f}\left[I_{2,m}-\varrho\frac{\partial\varepsilon_{m}}{\partial B}\right]+\sum_{m=f+1}^{\infty}I_{2,m},\label{Omegaoscall}
	\end{align}
	where $f$ is the last sorted position such that $\varepsilon_{f}\leq\mu<\varepsilon_{f+1}$,
	as considered in the previous section. Since we are interested only
	in the MO, we shall neglect the non-oscillatory contribution of the
	VB. In the appendix \ref{app:B} we show that if $\beta\mu\gg1$ and $\mu/\Gamma_{m}\gg1$, we
	can neglect the terms with $I_{1,m}$ and $I_{3,m}$.
	
	As done previously in the ground
	state, it is convenient to separate the MO in the indices $\gamma$.
	For each $\gamma$, the final position $f_{\gamma}$, determined by $\varepsilon_{f_{\gamma};\gamma}\leq\mu<\varepsilon_{f_{\gamma}+1;\gamma}$,
	can be obtained from the $\psi_{\gamma}$ function as $f_{\gamma}=\textrm{floor}\left(\psi_{\gamma}\right)=\left[\psi_{\gamma}\right]$
	(see equation (\ref{phase})). We define the occupancy factor
	$\mathcal{F}_{n;\gamma}$, for each energy level $\varepsilon_{n;\gamma}$,
	as
	\begin{equation}
		\mathcal{F}_{n;\gamma}=\int_{-\infty}^{\infty}\rho_{n;\gamma}\frac{1}{1+e^{\beta\left(\varepsilon-\mu\right)}}d\varepsilon.\label{Fit}
	\end{equation}
	At low damping the occupancy factor $\mathcal{F}_{n;\gamma}$ only
	changes (relative to its ground state value) when $\varepsilon_{n;\gamma}$
	is close to $\mu$. Thus it is good approximation to expand each $\varepsilon_{n;\gamma}$ around $n=\psi_{\gamma}$ and
	take $\left(\varrho/\mathcal{A}\right)\partial\varepsilon_{n;\gamma}/\partial B=-A_{\gamma}$,
	where $A_{\gamma}$ is the SO amplitude in the ground state, given
	by equation (\ref{Ans}). In this way, from equations (\ref{I2}) and (\ref{Omegaoscall})
	we obtain, for each $\gamma$, the MO with damping
	\begin{equation}
		M_{\textrm{osc},\gamma}=M_{\textrm{osc},\gamma}^{0}+A_{\gamma}\sum_{n\leq f_{\gamma}}\left(\mathcal{F}_{n;\gamma}-1\right)+A_{\gamma}\sum_{n>f_{\gamma}}\mathcal{F}_{n;\gamma}.\label{MgammaiT}
	\end{equation}
	The total magnetization is obtained by summing over all $\gamma$. 
	The MO, as expressed by equation (\ref{MgammaiT}), allows a simple interpretation.
	First of all, we should notice that in the ground state we have $\mathcal{F}_{n;\gamma}=1$ if $n\leq f_{\gamma}$
	and $\mathcal{F}_{n;\gamma}=0$ if $n>f_{\gamma}$, so $M_{\text{osc},\gamma}\rightarrow M_{\text{osc},\gamma}^{0}$
	as expected. At non zero damping, equation (\ref{MgammaiT})
	tell us that the MO are modified by the inclusion of the last two
	terms. Each one of these can be viewed as representing the change
	in the MO due to the modification of the available states. This is
	determined by the occupancy factor $\mathcal{F}_{n;\gamma}$ of each
	state $\varepsilon_{n;\gamma}$, and the corresponding modification
	to the MO, given by the oscillation amplitude $A_{\gamma}$.
	In other words, the states $n\leq f_{\gamma}$ (fully occupied in
	the ground state) are broadened and thermally excited, which tends
	to vacate them ($\mathcal{F}_{n;\gamma}<1$), changing the MO by a
	factor $A_{\gamma}\left(\mathcal{F}_{n;\gamma}-1\right)$. The states
	$n>f_{\gamma}$ (empty in the ground state) are also broadened and
	thermally excited, which tends to occupy them ($\mathcal{F}_{n;\gamma}>0$),
	changing the MO by a factor $A_{\gamma}\mathcal{F}_{n;\gamma}$. 
	
	Equation (\ref{MgammaiT}) can be conveniently rewritten in different
	ways, depending on the situation. For instance, if $\psi_{\gamma}\gg1$
	such that $M_{\text{osc},\gamma}^{0}$ is given by equation (\ref{Mosc high oc}),
	then one can use the properties of the arc tangent function to rewrite \cite{Escudero2018a}
	\begin{equation}
		M_{\text{osc},\gamma}=\frac{A_{\gamma}}{\pi}\arctan\left[\cot\left(\pi\psi_{\gamma}-\pi\sum_{n}\mathcal{F}_{n;\gamma}\right)\right].
	\end{equation}
	This expression would be particularly useful if one is interested
	in the overall profile of the MO, when many peaks are considered.
	On the other hand, when one analyses the damping effects
	over each MO peak, it is useful to rearrange $M_{\text{osc},\gamma}$ considering
	only the nearby states, because these are the ones that primarily
	modify the MO. For
	instance, around the magnetization peak at $\psi_{\gamma}=n_{0}$, it is good
	approximation to consider only the factors $\mathcal{F}_{n;\gamma}$
	close to $n=n_{0}$. In this way we can approximate
	\begin{align}
		M_{\text{osc},\gamma} \simeq & M_{\text{osc},\gamma}^{0}+A_{\gamma}\mathcal{F}_{n_{0};\gamma}\nonumber \\
		& +A_{\gamma}\left(\mathcal{F}_{n_{0}-1;\gamma}+\mathcal{F}_{n_{0}+1;\gamma}-1\right)+\ldots\label{Moscaprox}
	\end{align}
	The first order approximation is $M_{\text{osc},\gamma}\simeq M_{\text{osc},\gamma}^{0}+A_{\gamma}\mathcal{F}_{n_{0};\gamma}$,
	which is just the damping effects due only to the $n_{0}$ state.
	Around the $n_{0}$ magnetization peak, this is always the dominant
	term modifying the MO. Thus, at very low damping, the modification of the MO around the $n_{0}$ peak can be studied
	only by analysing the occupancy factor $\mathcal{F}_{n_{0};\gamma}$. 
	
	\subsection{Damping effects as reduction factors}\label{subsec:3.2}
	
	We shall show that, if the broadening is the same for all LL
	(so that $\Gamma$ does not depend on $n$), then the damping can be described with reduction factors.
	We start by rewriting equation (\ref{MgammaiT}) as 
	\begin{align}
		M_{\textrm{osc},\gamma}\left(\mu\right)  =& \int_{-\infty}^{\infty}\delta\left(\varepsilon-\mu\right)M_{\textrm{osc},\gamma}^{0}\left(\varepsilon\right)d\varepsilon-\sum_{n=1}^{\infty}\frac{\varrho}{\mathcal{A}}\frac{\partial\varepsilon_{n;\gamma}}{\partial B}\nonumber \\
		& \times\int_{-\infty}^{\infty}\left[\mathcal{F}_{\gamma}-H\left(\varepsilon-\mu\right)\right]\delta\left(\varepsilon-\varepsilon_{n;\gamma}\right)d\varepsilon,\label{Mr1}
	\end{align}
	where $H\left(\varepsilon-\mu\right)$ is the Heaviside function and 
	\begin{equation}
		\mathcal{F_{\gamma}}=\int_{-\infty}^{\infty}\rho_{\gamma}\left(\varepsilon'-\varepsilon\right)n_{F}\left(\varepsilon'-\mu\right)d\varepsilon',
	\end{equation}
	where $n_{\text{F}}=\left[1+e^{\beta\left(\varepsilon'-\mu\right)}\right]^{-1}$
	is the Fermi-Dirac distribution. Integrating by parts the second integral
	in equation (\ref{Mr1}) result
	\begin{align}
		M_{\textrm{osc},\gamma}\left(\mu\right)  =&\int_{-\infty}^{\infty}\delta\left(\varepsilon-\mu\right)M_{\textrm{osc},\gamma}^{0}\left(\varepsilon\right)d\varepsilon+\sum_{n=1}^{\infty}\frac{\varrho}{\mathcal{A}}\frac{\partial\varepsilon_{n;\gamma}}{\partial B}\nonumber \\
		&\times\int_{-\infty}^{\infty}\left[\mathcal{F}'_{\gamma}+\delta\left(\varepsilon-\mu\right)\right]H\left(\varepsilon-\varepsilon_{n;\gamma}\right)d\varepsilon,
	\end{align}
	where 
	\begin{equation}
		\mathcal{F}'_{\gamma}=\int_{-\infty}^{\infty}\rho_{\gamma}\left(\varepsilon'-\varepsilon\right)\frac{\partial n_{F}}{\partial\varepsilon'}\left(\varepsilon'-\mu\right)d\varepsilon'.\label{Mr2}
	\end{equation}
	Thus, from equation (\ref{M0}) we obtain
	\begin{equation}
		M_{\textrm{osc},\gamma}\left(\mu\right)=-\int_{-\infty}^{\infty}\rho_{\gamma}\left(\varepsilon'-\varepsilon\right)\left[-\frac{\partial n_{F}}{\partial\varepsilon'}\right]M_{\textrm{osc},\gamma}^{0}\left(\varepsilon\right)d\varepsilon d\varepsilon'
	\end{equation}
	Hence the damping effects are expressed as convolutions of the ground
	state magnetization, which is the basis of the LK formula. From this,
	the MO with the reduction factors can be obtained by replacing in
	equation (\ref{Mr2}) the SO of $M_{\textrm{osc},\gamma}^{0}\left(\varepsilon\right)$,
	given by equation (\ref{Mns}). This is done in the appendix \ref{app:C}, where
	it is shown that 
	\begin{equation}
		M_{\text{osc},\gamma}=A_{\gamma}\sum_{p=1}^{\infty}R_{T,\gamma}\left(p\right)R_{\Gamma,\gamma}\left(p\right)\frac{\sin\left(2\pi p\psi_{\gamma}\right)}{\pi p},\label{MLKgeneral}
	\end{equation}
	where $R_{T,\gamma}\left(p\right)$ and $R_{\Gamma,\gamma}\left(p\right)$
	are the temperature and broadening reduction factors, respectively,
	given by
	\begin{align}
		R_{T,\gamma}\left(p\right) =& \left(\frac{2\pi^{2}p}{\beta}\frac{\partial\psi_{\gamma}}{\partial\mu}\right)\textrm{csch}\left(\frac{2\pi^{2}p}{\beta}\frac{\partial\psi_{\gamma}}{\partial\mu}\right)\label{Rt}\\
		R_{\Gamma,\gamma}\left(p\right) =& \int_{-\infty}^{\infty}\rho_{\gamma}\left(x\right)\cos\left[2\pi px\frac{\partial\psi_{\gamma}}{\partial\mu}\right]dx.\label{Rg}
	\end{align}
	The MO, as given by equation (\ref{MLKgeneral}), is a generalization of
	the LK formula in 2D, for a general system with energy levels $\varepsilon_{n;\gamma}$.
	The total MO is obtained by summing over all $\gamma$. 
	The expression for $R_{T,\gamma}\left(p\right)$ is the known temperature reduction factor of the LK formula,
	generalized to be dependent on $\psi_{\gamma}$. The
	expression for $R_{\Gamma,\gamma}\left(p\right)$ depends on how is the Landau level broadening. In the appendix \ref{app:D} we obtain $R_{\Gamma}$ for some common DOS distributions. 
	
	\begin{figure*}
		\includegraphics[scale=0.39]{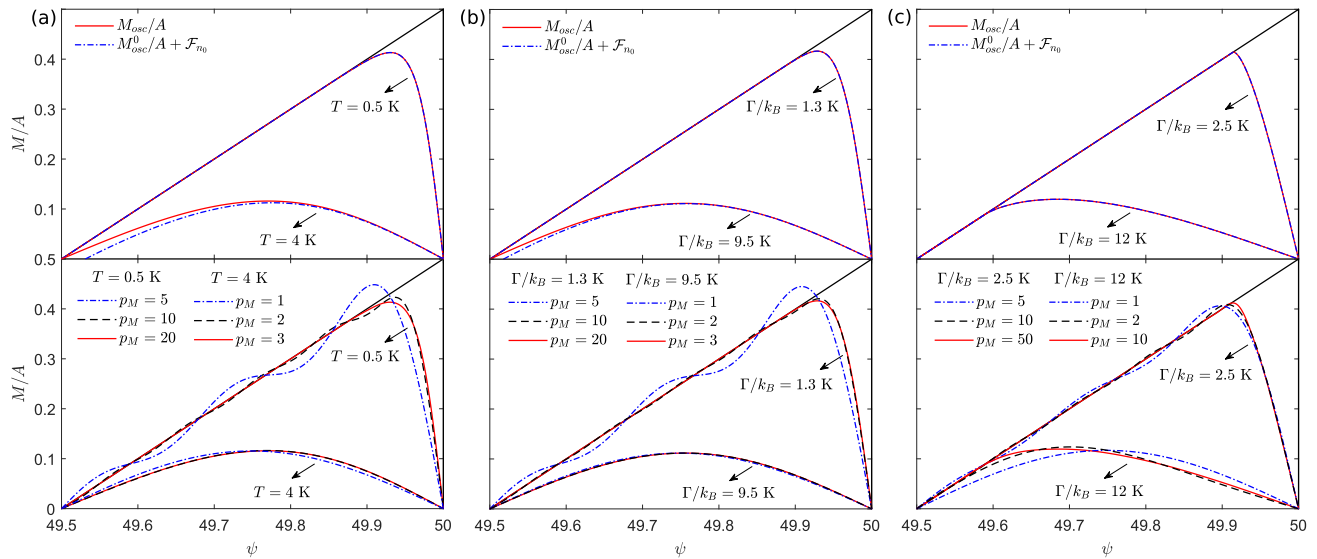}
		\caption{MO of a system with 2D relativistic Landau levels $\varepsilon_{n}=a\sqrt{Bn}$ with $\mu=0.25$ eV, as a function of the MO phase $\psi=\left(\mu/a\right)^{2}/B$, around the peak at $\psi=n_0=50$. In the top figures we plot the MO with the damping effects in terms of the Landau level occupancy. In the bottom figures we plot the MO with the damping effects as reduction factors. In both approaches we show the approximations given by the occupancy factor of the $n_0$ Landau level (top figures) and the sum of the first $p_M$ harmonic terms (bottom figures). We consider the damping cases of: (a) finite temperature and zero broadening; zero temperature and broadening $\Gamma$ with (b) gaussian and (c) semi-elliptic distribution. In each case we consider \textit{low} and \textit{high} damping situations, with the damping parameters chosen to represent similar MO amplitude reduction. In all cases it is also plotted, in black solid line, the ground state magnetic sawtooth oscillation.}\label{fig:1} 
	\end{figure*}
	
	Equation (\ref{MLKgeneral}) is particularly useful when the infinite
	sum can be solved, as it is the case in some special situations. For
	instance, for a Lorentz distribution, the broadening factor is the known Dingle factor $R_{\Gamma,\gamma}\left(p\right)=\exp\left(-2\pi p\Gamma_{\gamma}\partial\psi_{\gamma}/\partial\mu\right)$ (see appendix \ref{app:D}), for which at zero temperature the infinite sum given by equation (\ref{MLKgeneral}) can be solved. Likewise when the temperature is relatively high and $R_{T,\gamma}\left(p\right)$ can be approximated as a simple exponential.
	This is a very good approximation for the MO in metals \cite{Shoenberg1984}, but
	it does not hold at the very low temperatures required to observe
	the fine details of the MO in 2D materials \cite{Escudero2018a,Escudero2019}. 
	
	From equations (\ref{Rt}) and (\ref{Rg}) we can already see that the
	reduction factors depend on the energy levels through the term $\left(\partial\psi_{\gamma}/\partial\mu\right)$.
	For classical 2D LL we have $\psi_{\gamma}\sim\mu$, whereas
	for relativistic LL (like in graphene) we have $\psi_{\gamma}\sim\mu^{2}$ (see equation (\ref{phase})).
	Thus in the classical 2D electron gas, the reduction factors do not
	depend on the chemical potential, whereas in graphene they are proportional
	to $\mu$ \cite{Sharapov2004}.

	\subsection{Analysis of the two damping descriptions}\label{subsec:3.3}
	
	We shall compare and discuss the two methods of describing the damping
	effect in the MO, as developed in sections \ref{subsec:3.1} and \ref{subsec:3.2}, considering broadening
	distributions with constant $\Gamma$. As an example,
	we consider relativistic 2D Landau levels, which occur in graphene-like
	systems. For the sake of simplicity we ignore any effect of spin splitting
	or valley mixing. The energy levels are $\varepsilon_{n}=a\sqrt{Bn}$,
	where $a$ is material dependent constant. The corresponding phase
	of the ground state MO is $\psi=\left(\mu/a\right)^{2}/B$ (see equation (\ref{phase})). We consider
	$B$ and $\mu$ such that $\psi$ is relatively high, in which case $M_{\text{osc}}^{0}$ is given by equation (\ref{Mosc high oc}).
	
	We will consider the simple cases of (a) finite temperature and zero broadening, (b) zero temperature with Gaussian broadening and (c) zero temperature with semi-elliptic broadening. 
	In each case, the occupancy factor $\mathcal{F}_{n}$ and the reduction
	factors $R$ are obtained in the appendix \ref{app:D}. It should be noted that we are considering cases in which the infinite series of equation (\ref{MLKgeneral}) cannot be solved. Thus we do not consider, for instance, the case of broadening with a Lorentz distribution at zero temperature (see above discussion).  
	For each case,
	the MO are given by equations (\ref{MgammaiT}) and (\ref{MLKgeneral}).
	In general, each expression can be approximated. We consider a magnetic
	field $B$ such that $n_{0}-1<\psi<n_{0}$ so $\left[\psi\right]=n_{0}-1$,
	where $n_{0}\gg1$. Then for equation (\ref{MgammaiT}) we shall use the first
	order in equation (\ref{Moscaprox}), whereas in equation (\ref{MLKgeneral}) we take
	the upper limit $p_{M}$ in the summation. 
	
	In Figure \ref{fig:1} it is shown the MO as a function of $\psi$, with fixed $\mu=0.25$
	eV. For the three cases, we show the damping effects around the peak
	$n_{0}=50$, considering $n_{0}-1/2<\psi<n_{0}$. In each case we plot
	the SO $M_{\text{osc}}^{0}$ (black solid line) and the MO given by the LL occupancy approach (top
	figures) and the reduction factor approach (bottom figures), considering its approximations. The temperature $T$ and broadening parameter $\Gamma$ were specifically chosen to show a similar MO amplitude reduction, considering the two cases of \textit{low} and \textit{high} damping. As we can observe, the two ways to describe the MO with damping give the same result, as expected. Nevertheless, they differ substantially in how effective they are, particularly depending if the damping is high or low. 
	
	At low damping, we see that the occupancy factor description (top figures) can be done by just considering the occupancy factor of the $n_0$ LL. This can be very convenient in order to study how the increase of temperature or impurities affects the observation of the fine details in the MO of 2D materials, such as those due to the spin splitting or valley mixing. In other words, at low damping the broadening of any SO is entirely dictated by how is the occupancy factor of the LL that causes the SO peak. In contrast, at low damping the reduction factor description of the MO requires more terms. As can be seen in the bottom figures, at low damping one needs to consider many terms in the harmonic summation of equation (\ref{MLKgeneral}), in order to converge the series. This implies that one cannot describe the MO with just the first terms in the series, as is usually done when using the LK formula in metals.
	
	At higher damping we observe a different behaviour. In the top figures we see that the MO cannot be described by just the occupancy of the $n_0$ LL. This can be seen in the (a) and (b) cases, although it should be mentioned that the MO can still be perfectly described by the next term in the expansion given by equation (\ref{Moscaprox}). That is, for $T=4$ K and $\Gamma/k_{\text{B}}=9.5$ K (Gauss broadening), we have $M_{\textrm{osc}}/A=M_{\textrm{osc}}^{0}/A+\mathcal{F}_{n_{0}}+\left(\mathcal{F}_{n_{0}-1}+\mathcal{F}_{n_{0}+1}-1\right)$. Thus the difference between the first order approximation (blue dot-dashed line) and the total MO (red solid line) is caused by the change in the occupancy of the nearby $n_0-1$ and $n_0+1$ LL. For the (c) case we observe that even at relatively high damping, the MO depend on just the occupancy of the LL $n_0$. This is expected considering the nature of the semi-elliptic broadening, in which if $\Gamma<\Delta\varepsilon_{n}$ then the broadening of each level is independent. On the other hand, in the bottom figures we observe that at \textit{high} damping the reduction factor description requires fewer harmonic terms. This can become convenient, especially when the reduction factors can be approximated so that the summation in equation (\ref{MLKgeneral}) can be solved.
	
	\subsection{Comparison between different broadening distributions}\label{subsec:3.4}
	
\begin{figure}
		\includegraphics[scale=0.39]{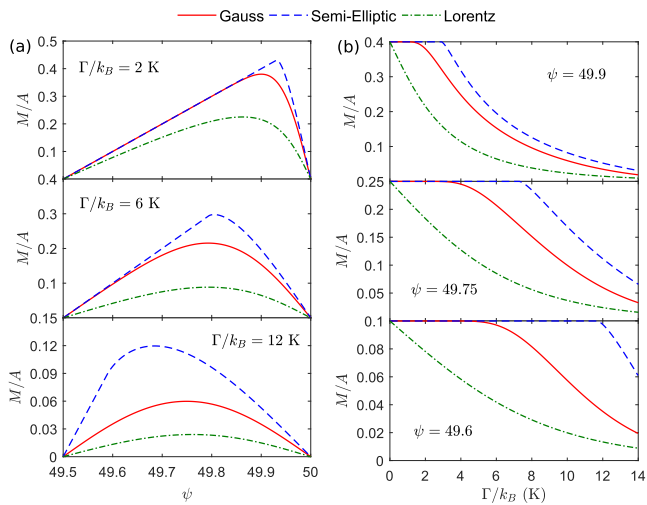}
		\caption{MO of a system with 2D relativistic Landau levels $\varepsilon_{n}=a\sqrt{Bn}$ with $\mu=0.25$ eV, for different broadening distributions at zero temperature. In (a) we plot the MO in terms of the MO phase $\psi=\left(\mu/a\right)^{2}/B$, around the peak at $\psi=50$, considering different broadening parameters $\Gamma$. In (b) we plot the MO as function of the $\Gamma$, at different phases $\psi$.}\label{fig:2}  
	\end{figure}

	The damping of the MO can depend significantly on the broadening of the LL. This not only depends on the type of broadening distribution, but also on the broadening parameter $\Gamma$ itself. In Figure \ref{fig:2} we show the MO of a system with 2D relativistic Landau levels $\varepsilon_{n}=a\sqrt{Bn}$ with $\mu=0.25$ eV (as in Figure \ref{fig:1}), at zero temperature and for broadening of type Gauss (red solid line), semi-elliptic (blue dashed line) and Lorentz (green dot-dashed line). Specifically, in Figure \ref{fig:2}(a) we show the MO as a function of the phase $\psi=\left(\mu/a\right)^{2}/B$, around the peak at $\psi=n_0=50$, considering different broadening parameters $\Gamma$. In Figure \ref{fig:2}(b) we show the MO as a function of the $\Gamma$, at different fixed phases $\psi$ (which implies constant magnetic field). 
	
	As we can see in Figure \ref{fig:2}(a), the shape of the MO with damping parameter $\Gamma$ depends strongly on the type of broadening distribution. The Lorentz distribution decays always faster, whereas the semi-elliptic distribution decays always slower. At low damping, for both the Gauss and the semi-elliptic distributions, the MO are not damped if the phase is not close to $n_0=50$. In other words, further from $\psi=50$, the MO reduce to the ground state magnetization (sawtooth oscillation in this case). The same behaviour was also seen in Figure \ref{fig:1}, where at low damping the MO eventually reduces to the ground state MO (black solid line). The reason is that at low damping, both the Gauss and semi-elliptic distributions decay fast and thus $\mathcal{F}_{n_0}\ll1$ unless $\psi$ is close to $n_0$. This is not the case for the Lorentz distribution, due to its long range which implies that $\mathcal{F}_{n_0}$ decays much slower. 
	
	The dependence of the MO decay with the type of broadening can be also seen in Fig 2(b). We observe that the decay of the MO depends on how far is the phase $\psi$ from $n_0$, at which the energy level cross the Fermi energy. In the three considered $\psi$, the top value in the figure corresponds to the value of the ground state magnetization $M/A=\arctan\left[\cot\left(\pi\psi\right)\right]/\pi$,
	which for $49.5<\psi<50$ implies $M/A=\psi-49.5$. Hence, for the Gauss and semi-elliptic distributions, we see that the further $\psi$ is from $n_0$ (for $49.5<\psi<50$), the larger is the broadening $\Gamma$ required to damp the MO. For instance, at $\psi=49.75$ we observe the MO are damped only if $\Gamma/k_{\text{B}}>3$ K for the Gauss distribution and $\Gamma/k_{\text{B}} >7$ K for the semi-elliptic distribution. In contrast, for the Lorentz distribution the MO are practically always damped, regardless of $\psi$ and $\Gamma$.
	
	\subsection{Chemical potential oscillations}\label{subsec:3.5}
	
	The presented formulation assumes a constant chemical potential $\mu$, which necessarily implies that the number of electrons is not constant as the magnetic field is varied.
	The satisfaction of this condition depends on the configuration
	of the system, such as the application of a gate voltage.
	Nevertheless, in most cases it is the electron density $n_{e}=N/\mathcal{A}$
	of the system that is kept constant. In that situation one has to
	obtain how should $\mu$ oscillate as a function of $B$ in order
	to maintain $n_{e}$ constant \cite{Grigoriev2001,Champel2001}. In three dimensional (3D) metals, the
	oscillation in $\mu$ is very small and usually can be neglected
	\cite{Shoenberg1984}. In 2D systems, these oscillations
	can be appreciable and thus one needs to be careful about which situation
	(constant chemical potential or electron density) best represents
	the system \cite{Shoenberg1984a}. The oscillations of $\mu$ are obtained from $N=-\left(\partial\Omega/\partial\mu\right)$,
	which using equations (\ref{Om}) and (\ref{Fit}), can be written as
	\begin{equation}
		N=\varrho\sum_{\gamma}\sum_{n}\mathcal{F}_{n;\gamma}.
	\end{equation}
	From this equation one should solve $\mu=\mu\left(n_{e},B,\beta,\Gamma\right)$, which in
	general can only be done numerically, although at low damping it is possible to make some simplifications. For instance, it is known that in the ground
	state, $\mu$ reduces to the last partially occupied energy level \cite{Shoenberg1984a}.
	Thus at low damping it is good approximation to consider only the
	$\mu$ dependence in the $\mathcal{F}_{n;\gamma}$ with $\varepsilon_{n;\gamma}$
	close to the last occupied energy level, considering for the rest
	$\mathcal{F}_{n;\gamma}$ their ground state value (1 or 0). 
	
	\section{Conclusions}\label{sec:4}
	
	We developed a general formalism to obtain the magnetic oscillations (MO) in two dimensional (2D) systems. We considered a general expression for the 2D Landau levels (LL), which may depend not only on the perpendicular magnetic field that causes them, but also on other variables and indices. This may included perpendicular and parallel electric fields, or the spin, valley and wave vector indices. In the ground state of the system, we obtained expressions for the MO amplitude and phase. Using this and a Fourier expansion, we wrote the MO as a sum of elemental oscillations. The first term is a sawtooth oscillation, which becomes dominant when many energy levels are occupied.
	
	The effects of temperature, impurities or lattice imperfections were also considered. We assumed a general density of states that broaden each energy level, taking into account that the broadening parameter may depend on the magnetic field and the LL index. From this we obtained an expression for the damping effects in the MO, which depends on the occupancy of the LL. We then showed that, if the broadening parameter does not depend on the LL, the damping effects can also be described using reduction factors, which resulted in a generalization of the known Lifshits-Kosevich (LK) formula. We compared both approaches, for graphene-like LL and different types of damping effects. At low damping, we showed that the MO depend only on the occupancy factor of the LL closest to the Fermi energy, whereas at higher damping one needs to consider also the occupancy of the nearby levels. In contrast, the lower the damping the more harmonic terms one needs to consider in the LK formula in order to converge the series; at higher damping, only the first few harmonic are needed.
	
	The general formulation presented gives a relatively direct way to obtain the MO in 2D systems, by calculating its more important properties: the amplitude and phase of the oscillations. The expressions obtained show that the MO is mostly dependent on the properties of the energy levels. Hence, by measuring the MO profile, one could map the corresponding LL properties. The general description of the damping effects in the MO can be useful to study how is the scattering in 2D materials under a perpendicular magnetic field. In particular, how are the LL broadened. Careful magnetization measurements could be one of the most reliable methods, as there are no perturbations other than the magnetic field that causes the LL. 
	
	\begin{acknowledgements}
		
		This paper was partially supported by grants of CONICET (Argentina
		National Research Council) and Universidad Nacional del Sur (UNS)
		and by ANPCyT through PICT 2014-1351. Res. N 270/15. N: 2014-1351, and PIP 2014-2016. Res. N 5013/14. C\'odigo: 11220130100436CO research grant, as well as by SGCyT-UNS., J. S. A. and P. J. are members of CONICET, F. E. acknowledge  research fellowship from this institution.
		
	\end{acknowledgements}
	
	\appendix
	\section{Fourier expansion of the ground state MO}\label{app:A}
	
	Written in terms of the phase $\psi_{\gamma}$ defined in equation (\ref{phase}),
	the ground state magnetization $M_{\gamma}^0$ oscillates as a function of $\psi_{\gamma}$
	in integer intervals. The oscillation is determined by the change
	in the last LL occupied when $\psi_{\gamma}$ is an integer and thus
	$\textrm{floor}\left(\psi_{\gamma}\right)$ changes. Considering that in general $\varepsilon_{n;\gamma}$
	increases with $n$ and $B$, then $\psi_{\gamma}$ decreases with increasing
	$B$. 
	This implies that in the small-field limit, when many energy levels
	are occupied and $\psi_{\gamma}$ is large, the magnetization varies
	little over each interval $\left(n_{0},n_{0}+1\right)$, where $n_{0}=\textrm{floor}\left(\psi_{\gamma}\right)$.
	Thus for any $x_{0}\in\left(n_{0},n_{0}+1\right)$ we can take the
	quadratic approximation
	\begin{align}
		M_{\gamma}^{0}&\simeq M_{\gamma}^{0}\left(x_{0}\right)+\frac{\partial M_{\gamma}^{0}}{\partial\psi_{\gamma}}\left(x_{0}\right)\left(\psi-x_{0}\right)\nonumber\\
		&\quad+\frac{\partial^{2}M_{\gamma}^{0}}{\partial\psi_{\gamma}^{2}}\left(x_{0}\right)\frac{\left(\psi-x_{0}\right)^{2}}{2}.\label{M taylor}
	\end{align}
	Using this approximation, the oscillating part of $M_{\gamma}^{0}$ can be expanded in a Fourier
	series. Because $M_{\gamma}^{0}$ oscillates discontinuously, the
	linear term gives a sawtooth oscillation, whereas the quadratic term
	gives a $\cos\left(2\pi p\psi\right)/\left(\pi p\right)^{2}$ term.
	That is
	\begin{equation}
		M_{\textrm{osc},\gamma}^{0}\sim C_{1}\sum_{p=1}^{\infty}\frac{\sin\left(2\pi p\psi\right)}{\pi p}+C_{2}\sum_{p=1}^{\infty}\frac{\cos\left(2\pi p\psi\right)}{\left(\pi p\right)^{2}},\label{M fourier}
	\end{equation}
	where $C_{1}$ and $C_{2}$ are $p$ independent factors. In this
	way, the oscillation part of the magnetization is written as the sum
	of elemental oscillating functions. The expansion given by equation (\ref{M fourier})
	is in general valid for any oscillating interval $\left(n_{0},n_{0}+1\right)$,
	as long as the small-field limit condition is satisfied. Hence the
	equation (\ref{M fourier}) can be generalized to all $\psi$ by considering
	the general dependence of the coefficients $C_{1}$ and $C_{2}$ with
	$\psi$. For $C_{1}$ we already have $C_{1}\rightarrow A_{\gamma}$,
	where $A_{\gamma}$ is the amplitude of the discontinuities in $M_{\gamma}$,
	given by equation (\ref{Ans}). It is worth noting that, if necessary, the next terms
	in the expansion (\ref{M fourier}) can be found by considering higher
	orders in the approximation given by equation (\ref{M taylor}) and expanding
	each one in a Fourier series. This results in elemental oscillations of
	the form $\sin\left(2\pi p\psi\right)/\left(\pi p\right)^{2n+1}$
	and $\cos\left(2\pi p\psi\right)/\left(\pi p\right)^{2n}$, where
	$n$ is a positive integer. However, in general the need of these
	terms would be required only in the extreme situation of high magnetic
	fields and very low occupancy. Finally, it should be mentioned that a similar expansion
	can be made for the grand potential $\Omega_{\textrm{osc},\gamma}$,
	only that then there is no sawtooth oscillation because $\Omega_{\textrm{osc},\gamma}$
	varies continuously and thus $C_{1}=0$.
	
	\section{Approximation of the MO for $\beta\mu\gg1$ and $\mu/\Gamma\gg1$}\label{app:B}
	
	We will show that if $\beta\mu\gg1$ and $\mu/\Gamma\gg1$,
	then we can neglect the terms with $I_{1,m}$ and $I_{3,m}$ in the oscillating
	part of equation (\ref{Omegaoscall}). We start with the separation of
	the terms for $m\leq f$ and $m\geq f+1$, as done in equation (\ref{Omegaoscall}).
	We define $I_{i,>}=\sum_{m=f+1}^{\infty}I_{i,m}$ for $i=\left\{ 1,2,3\right\} $,
	and 
	\begin{align}
		I_{1,<}  &= \sum_{m=1}^{f}\left[I_{1,m}-\frac{\partial\varrho}{\partial B}\left(\varepsilon_{m}-\mu\right)\right]\\
		I_{2,<}  &=   \sum_{m=1}^{f}\left[I_{2,m}-\varrho\frac{\partial\varepsilon_{m}}{\partial B}\right]\\
		I_{3,<}  &= \sum_{m=1}^{f}I_{3,m}.
	\end{align}
	The idea is to show that, if $\beta\mu\gg1$ and $\mu/\Gamma\gg1$, the oscillating amplitudes
	of $I_{1}$ and $I_{3}$ are much smaller than the oscillating amplitude
	of $I_{2}$. In each case, this amplitude is determined when $f$
	changes, which happens when the energy levels $\varepsilon_{m}$ cross
	the chemical potential $\mu$. Suppose $f$ changes to $f+1$ at $B=B_{f}$,
	such that $\varepsilon_{f+1}\left(B_{f}\right)=\mu$. Then the corresponding
	amplitudes are
	\begin{align}
		\Delta I_{1,<} &= I_{1,f+1}=-\Delta I_{1,>}\label{DeltaI1}\\
		\Delta I_{2,<} &=  I_{2,f+1}-\varrho\frac{\partial\varepsilon_{f+1}}{\partial B},\quad\Delta I_{2,>}=-I_{2,f+1}\label{DeltaI2}\\
		\Delta I_{3,<} &= I_{3,f+1}=-\Delta I_{3,>},\label{DeltaI3}
	\end{align}
	where, defining $y=\left(\varepsilon-\varepsilon_{f+1}\right)=\left(\varepsilon-\mu\right)$,
	from equations (\ref{I1}), (\ref{I2}) and (\ref{I3}) we can write
	\begin{align}
		I_{1,f+1} & = -\frac{1}{\beta}\frac{\partial\varrho}{\partial B}\int_{-\infty}^{\infty}\rho_{f+1}\ln\left[1+e^{\beta y}\right]dy\label{I1f}\\
		I_{2,f+1} & = \varrho\frac{\partial\varepsilon_{f+1}}{\partial B}\int_{-\infty}^{\infty}\rho_{f+1}\frac{1}{1+e^{\beta y}}dy\label{I2f}\\
		I_{3,f+1} & = -\frac{\varrho}{\beta}\frac{\partial\Gamma_{f+1}}{\partial B}\int_{-\infty}^{\infty}\frac{\partial\rho_{f+1}}{\partial\Gamma}\ln\left[1+e^{\beta x}\right]dy.\label{I3f}
	\end{align}
	Now, given that $\int_{-\infty}^{\infty}\rho_{f+1}dy=1$, we have
	\begin{align}
		\int_{-\infty}^{\infty}\rho_{f+1}dy = & \int_{-\infty}^{\infty}\rho_{f+1}\left[\frac{1}{1+e^{\beta y}}+\frac{1}{1+e^{-\beta y}}\right]dy\nonumber \\
		= & 2\int_{-\infty}^{\infty}\rho_{f+1}\frac{1}{1+e^{\beta y}}dx=1.\label{Inthalf}
	\end{align}
	Then we get $I_{2,f+1}=\varrho\left(\partial\varepsilon_{f+1}/\partial B\right)/2$,
	and in order of magnitude $\left(\partial\varepsilon_{f+1}/\partial B\right)\sim\mu/B$.
	Thus from equation (\ref{I2f}) we obtain
	\begin{equation}
		\left|\Delta I_{2,<}\right|=\left|\Delta I_{2,>}\right|\equiv\left|\Delta I_{2}\right|\sim\varrho\frac{\mu}{B},\label{I2oder}
	\end{equation}
	independently of the damping parameters $T$ and $\Gamma$. This is
	the same order as the sawtooth amplitude in the ground state (see
	equation (\ref{Ans})), which explains why the damped MO falls to zero
	when the energy levels cross $\mu$. On the other hand, for $I_{1}$
	and $I_{3}$, the order of magnitude of the oscillating amplitude
	depends on the damping parameters $T$ and $\Gamma$. This can be
	seen by inspection of equations (\ref{I1f}) and (\ref{I3f}), whose value
	depends on $T$ and $\Gamma$. For instance, without broadening such
	that $\rho_{f+1}\left(y\right)\rightarrow\delta\left(y\right)$, one
	has $\left|I_{1,f+1}\right|\sim\varrho/B\beta$ ($\partial\varrho/\partial B=\varrho/B$
	if $\varrho$ depends on $B$) and $I_{3,f+1}=0$. Likewise, at zero temperature, one has $\left|I_{1,f+1}\right|\sim\varrho\Gamma/B$
	and also $\left|I_{3,f+1}\right|\sim\varrho\Gamma/B$. In the general
	case, for any temperature and broadening, we can say $\left|I_{1,f+1}\right|\sim\varrho\left\{ \Gamma\right\} _{1}/B$
	and $\left|I_{1,m}\right|\sim\varrho\left\{ \Gamma\right\} _{2}/B$,
	where $\left\{ \Gamma\right\} _{1}$ and $\left\{ \Gamma\right\} _{3}$
	are terms that depend on the damping parameters $k_{\text{B}}T$ and $\Gamma$.
	We are not interested exactly in how are these terms, but rather in
	the fact that $\left\{ \Gamma\right\} _{1}$ and $\left\{ \Gamma\right\} _{3}$
	will always increase if $\Gamma$ and $T$ increase. Then, from equations
	(\ref{DeltaI1}) and (\ref{DeltaI3}) we can say
	\begin{align}
		\left|\Delta I_{1,<}\right|= &\left|\Delta I_{2,>}\right|\equiv\left|\Delta I_{1}\right|\sim\varrho\frac{\left\{ \Gamma\right\} _{1}}{B}\label{I1oder}\\
		\left|\Delta I_{3,<}\right|= &\left|\Delta I_{3,>}\right|\equiv\left|\Delta I_{3}\right|\sim\varrho\frac{\left\{ \Gamma\right\} _{3}}{B}.\label{I3oder}
	\end{align}
	From equations (\ref{I2oder}), (\ref{I1oder}) and (\ref{I3oder}) we
	obtain the ratios
	\begin{align}
		\left|\frac{\Delta I_{1}}{\Delta I_{2}}\right| \sim & \frac{\left\{ \Gamma\right\} _{1}}{\mu}\\
		\left|\frac{\Delta I_{3}}{\Delta I_{2}}\right| \sim & \frac{\left\{ \Gamma\right\} _{3}}{\mu}.
	\end{align}
	Therefore, under the condition $\beta\mu\gg1$ and $\mu/\Gamma\gg1$,
	we have $\left|\Delta I_{1}/\Delta I_{2}\right|\ll1$ and $\left|\Delta I_{3}/\Delta I_{2}\right|\ll1$,
	so the terms $I_{1,m}$ and $I_{3,m}$ can be neglected in the oscillating
	part of equation (\ref{Omegaoscall}). It is worth noting this low damping condition is usually required
	in order to observe the MO.
	
	\section{MO with damping effects as reduction factors}\label{app:C}
	
	We will obtain, from equation (\ref{Mr2}), the MO with the damping effects
	as reduction factors. We start by taking the SO of $M_{\gamma}^{0}\left(\varepsilon\right)$,
	given by equation (\ref{Mns}). We have
	\begin{equation}
		M_{\text{osc},\gamma}^0\left(\varepsilon\right)=A_{\gamma}\left(\varepsilon\right)\sum_{p=1}^{\infty}\frac{1}{\pi p}\sin\left[2\pi p\psi_{\gamma}\left(\varepsilon\right)\right],\label{Me}
	\end{equation}
	where we explicitly indicated that $A_{\gamma}$ and $\psi_{\gamma}$
	are obtained considering a chemical potential equal to $\varepsilon$.
	Replacing in equation (\ref{Mr2}) we get
	\begin{equation}
		M_{\text{osc},\gamma}\left(\mu\right)=\int_{-\infty}^{\infty}\left[-\frac{\partial n_{F}}{\partial\varepsilon'}\left(\varepsilon'-\mu\right)\right]Y_{\gamma}\left(\varepsilon'\right)d\varepsilon',\label{Mr3}
	\end{equation}
	where
	\begin{align}
		Y_{\gamma}\left(\varepsilon'\right) = & \int_{-\infty}^{\infty}\rho\left(\varepsilon'-\varepsilon\right)A_{\gamma}\left(\varepsilon\right)\sum_{p=1}^{\infty}\frac{\sin\left[2\pi p\psi_{\gamma}\left(\varepsilon\right)\right]}{\pi p}d\varepsilon\nonumber \\
		= & \sum_{p=1}^{\infty}\int_{-\infty}^{\infty}\rho\left(x\right)A_{\gamma}\left(\varepsilon'-x\right)\frac{\sin\left[2\pi p\psi_{\gamma}\left(\varepsilon'-x\right)\right]}{\pi p}dx.
	\end{align}
	At relatively low damping, the DOS $\rho\left(x\right)$ is non zero only for
	$\left|x\right|\ll1$, so is good approximation to take $A_{\gamma}\left(\varepsilon'-x\right)\simeq A_{\gamma}\left(\varepsilon'\right)$
	outside the integral. We can also take $\psi_{\gamma}\left(\varepsilon'-x\right)-\psi_{\gamma}\left(\varepsilon'\right)\simeq\left(\partial\psi_{\gamma}/\partial\varepsilon'\right)x$,
	so the sine function can be decomposed as
	\begin{align}
		\sin\left[2\pi p\psi_{\gamma}\left(\varepsilon'-x\right)\right] = & \sin\left[2\pi p\psi_{\gamma}\left(\varepsilon'\right)\right]\cos\left[2\pi p\frac{\partial\psi_{\gamma}}{\partial\varepsilon'}x\right]\nonumber \\
		& +\cos\left[2\pi p\psi_{\gamma}\left(\varepsilon'\right)\right]\sin\left[2\pi p\frac{\partial\psi_{\gamma}}{\partial\varepsilon'}x\right].\label{sine}
	\end{align}
	Then, given that $\rho_{\gamma}\left(x\right)$ is a symmetric function
	(as assumed), we have
	\begin{align}
		Y_{\gamma}\left(\varepsilon'\right) = & A_{\gamma}\left(\varepsilon'\right)\sum_{p=1}^{\infty}\frac{1}{\pi p}\sin\left[2\pi p\psi_{\gamma}\left(\varepsilon'\right)\right]\nonumber \\
		& \times\int_{-\infty}^{\infty}\rho\left(x\right)\cos\left[2\pi p\frac{\partial\psi_{\gamma}}{\partial\varepsilon'}x\right]dx.
	\end{align}
	From this we can identify the broadening reduction factor $R_{\Gamma}$,
	given by the integral in the last equation. That is, 
	\begin{equation}
		R_{\Gamma,\gamma}\left(p,\varepsilon'\right)=\int_{-\infty}^{\infty}\rho\left(x\right)\cos\left[2\pi p\frac{\partial\psi_{\gamma}}{\partial\varepsilon'}x\right]dx.
	\end{equation}
	Replacing in equation (\ref{Mr3}) and changing variables we obtain
	\begin{align}
		M_{\text{osc},\gamma}\left(\mu\right) = & \int_{-\infty}^{\infty}\left[-\frac{\partial n_{F}}{\partial x}\left(x\right)\right]A_{\gamma}\left(x+\mu\right)\nonumber \\
		& \times\sum_{p=1}^{\infty}\frac{R_{\Gamma,\gamma}\left(p,x+\mu\right)}{\pi p}\sin\left[2\pi p\psi_{\gamma}\left(x+\mu\right)\right]dx.\label{Mr4}
	\end{align}
	At low temperature, the function $\left(\partial n_{F}/\partial x\right)$
	is non zero only for $\left|x\right|\ll1$, so we can take $A_{\gamma}\left(x+\mu\right)\simeq A_{\gamma}\left(\mu\right)$
	and $R_{\Gamma,\gamma}\left(p,x+\mu\right)\simeq R_{\Gamma,\gamma}\left(p,\mu\right)$
	outside the integral. Then, decomposing again the sine function as
	in equation (\ref{sine}), and considering that $\left(\partial n_{F}/\partial x\right)$
	is a symmetric function, equation (\ref{Mr4}) becomes
	\begin{align}
		M_{\text{osc},\gamma}\left(\mu\right) = & A_{\gamma}\left(\mu\right)\sum_{p=1}^{\infty}\frac{R_{\Gamma,\gamma}\left(p,\mu\right)}{\pi p}\sin\left[2\pi p\psi_{\gamma}\left(\mu\right)\right]\nonumber \\
		& \times\int_{-\infty}^{\infty}\left[-\frac{\partial n_{F}}{\partial x}\right]\cos\left[2\pi p\frac{\partial\psi_{\gamma}}{\partial\mu}x\right]dx.
	\end{align}
	Hence we can identify the temperature reduction factor $R_{T}$, given
	by 
	\begin{equation}
		R_{T,\gamma}\left(p,\mu\right)=\int_{-\infty}^{\infty}\left[-\frac{\partial n_{F}}{\partial x}\left(x\right)\right]\cos\left[2\pi p\frac{\partial\psi_{\gamma}}{\partial\mu}x\right]dx.
	\end{equation}
	In this way, we finally obtain 
	\begin{equation}
		M_{\text{osc},\gamma}\left(\mu\right)=A_{\gamma}\sum_{p=1}^{\infty}R_{\Gamma,\gamma}\left(p,\mu\right)R_{T,\gamma}\left(p,\mu\right)\frac{\sin\left[2\pi p\psi_{\gamma}\left(\mu\right)\right]}{\pi p}.
	\end{equation}
	Thus the damping effects are described as reduction factors in the MO amplitude. 
	
	\section{Examples of damping effects in the MO}\label{app:D}
	
	Here we obtain the damping effects in the MO, as developed in sections \ref{subsec:3.1} and \ref{subsec:3.2}, for some particular cases. We do not particularize to any
	system, so the result correspond to any general 2D energy levels $\varepsilon_{n;\gamma}$.
	In each situation we obtain the occupancy factor $\mathcal{F}_{n;\gamma}$
	and the reduction factor $R_{\Gamma,\gamma}\left(p\right)$, given
	by equations (\ref{Fit}) and (\ref{Rg}), respectively. The temperature
	reduction factor $R_{T,\gamma}\left(p\right)$ has always the same
	general expression, given by equation (\ref{Rt}).
	
	\subsection{The pristine case}
	
	In the pristine case, the DOS is just the Dirac delta $\rho_{n;\gamma}=\delta\left(\varepsilon-\varepsilon_{n;\gamma}\right)$.
	Then the occupancy factor simply becomes
	\begin{equation}
		\mathcal{F}_{n;\gamma}=\frac{1}{1+e^{\beta\left(\varepsilon_{n;\gamma}-\mu\right)}},
	\end{equation}
	which is just the Fermi-Dirac distribution for the energy level $\varepsilon_{n;\gamma}$ \cite{Escudero2018a,Escudero2019}.
	For the reduction factors, at $\Gamma=0$ we have $R_{\Gamma,\gamma}\left(p\right)=1$,
	while the temperature reduction factor is given by equation (\ref{Rt}).
	
	\subsection{Non zero broadening at zero temperature}
	
	At zero temperature, $\beta\rightarrow\infty$ so equation (\ref{Fit})
	becomes
	\begin{equation}
		\mathcal{F}_{n;\gamma}=\int_{-\infty}^{\mu}\rho_{n;\gamma}d\varepsilon.\label{Fnzerot-1}
	\end{equation}
	To proceed one needs specify how is the broadening of the levels,
	that is, how is $\rho_{n;\gamma}$. Some common cases are the Lorentz
	(L), Gauss (G) and semi-elliptic (E) distributions, which have the
	form
	\begin{align}
		\textrm{(L)} \quad  \rho_{n;\gamma}=&\frac{\Gamma_{n;\gamma}}{\pi\left[\left(\varepsilon-\varepsilon_{n;\gamma}\right)^{2}+\Gamma_{n;\gamma}^{2}\right]}\\
		\textrm{(G)} \quad  \rho_{n;\gamma}=&\frac{1}{\Gamma_{n;\gamma}\sqrt{\pi}}e^{-\left(\varepsilon-\varepsilon_{n;\gamma}\right)^{2}/\Gamma_{n;\gamma}^{2}}\\
		\textrm{(E)} \quad  \rho_{n;\gamma}=&\frac{2}{\pi\Gamma_{n;\gamma}^{2}}\textrm{Re}\left[\sqrt{\Gamma_{n;\gamma}^{2}-\left(\varepsilon-\varepsilon_{n;\gamma}\right)^{2}}\right],
	\end{align}
	where $\Gamma_{n;\gamma}$ is the broadening parameter
	which in general can depend on the Landau level and other variables
	such as the magnetic field $B$. Then equation (\ref{Fnzerot-1}) gives
	\begin{align}
		\textrm{(L)} \quad  \mathcal{F}_{n;\gamma}=&\frac{1}{\pi}\left[\arctan\left(\frac{\mu-\varepsilon_{n;\gamma}}{\Gamma_{n;\gamma}}\right)+\frac{\pi}{2}\right]\\
		\textrm{(G)} \quad  \mathcal{F}_{n;\gamma}=&\frac{1}{2}\left[\textrm{erf}\left(\frac{\mu-\varepsilon_{n;\gamma}}{\Gamma_{n;\gamma}}\right)+1\right]\\
		\textrm{(E)} \quad  \mathcal{F}_{n;\gamma}=&\frac{1}{\pi}\left[\arctan\left(\frac{\mu-\varepsilon_{n;\gamma}}{\textrm{Re}\left[\sqrt{\Gamma_{n;\gamma}^{2}-\left(\mu-\varepsilon_{n;\gamma}\right)^{2}}\right]}\right)+\frac{\pi}{2}\right]\nonumber \\
		& +\frac{\left(\mu-\varepsilon_{n;\gamma}\right)}{\pi\Gamma_{n;\gamma}^{2}}\textrm{Re}\left[\sqrt{\Gamma_{n;\gamma}^{2}-\left(\mu-\varepsilon_{n;\gamma}\right)^{2}}\right].
	\end{align}
	For the reduction factors, at zero temperature $R_{T,\gamma}\left(p\right)=1$,
	while for $R_{\Gamma,\gamma}\left(p\right)$ one has to consider the
	broadening to be the same for all the Landau levels, so $\Gamma_{n;\gamma}=\Gamma_{\gamma}.$
	From equation (\ref{Rg}) we get
	\begin{align}
		\textrm{(L)} \quad  R_{\Gamma,\gamma}\left(p\right)=&\exp\left(-2\pi p\Gamma_{\gamma}\frac{\partial\psi_{\gamma}}{\partial\mu}\right)\\
		\textrm{(G)} \quad  R_{\Gamma,\gamma}\left(p\right)=&\exp\left[-\left(\pi p\Gamma_{\gamma}\frac{\partial\psi_{\gamma}}{\partial\mu}\right)^{2}\right]\\
		\textrm{(E)}  \quad R_{\Gamma,\gamma}\left(p\right)=&\frac{1}{\pi p\Gamma_{\gamma}\left(\partial\psi_{\gamma}/\partial\mu\right)}J_{1}\left(2\pi p\Gamma_{\gamma}\frac{\partial\psi_{\gamma}}{\partial\mu}\right),
	\end{align}
	where $J_{1}$ is the Bessel function of the first kind.
	
	\bibliography{references}	
	
\end{document}